\documentclass{ws-p9-75x6-50}

\begin{document}

\title{Luminosities of Disk--accreting Non--magnetic Neutron Stars}

\author{Arun V. Thampan}

\address{Inter--University Centre for Astronomy \& Astrophysics
(IUCAA), Pune 411 007, INDIA \\E-mail: arun@iucaa.ernet.in}

%


\maketitle

\abstracts{Disk accretion onto a neutron star possessing a
                      weak surface magnetic field ($B \le 10^8$~G)
                      provides interesting X-ray emission scenarios,
                      and is relevant for understanding X-ray
                      bursters and low-mass X-ray binaries (LMXBs). The
                      standard (Newtonian) theory of disk-accretion
                      predicts that the matter spiralling in from
                      infinity loses one-half of its total
                      gravitational energy in the extended disk, and
                      the remainder in a narrow boundary layer
                      girdling the neutron star.  The ratio of the
		      boundary layer luminosity to that from the
		      disk 
		      ($E_{\rm BL}/E_{\rm D}$) 
		      is, therefore, 
		      unity. On incorporation of general relativity 
		      without rotation (Schwarzschild solution), 
		      $E_{\rm BL}/E_{\rm D}$ is seen to be as high as
		      6.  We construct rotating sequences of neutron 
		      stars for three representative equations of
		      state.  We show here that for a neutron star 
		      rotating at a limit where centrifugal force 
		      balances the inward gravitational force, 
		      $E_{\rm BL}/E_{\rm D}\sim 0$.}

\section{Introduction}
Low mass X--ray binaries (LMXBs) comprise either a neutron star or a black 
hole, revolving around and pulling in matter from the outer envelopes
of, a binary companion star (M-K spectral type) that fills its Roche--lobe.
The accretion takes place via a disk and in the case of a neutron star, 
the angular momentum of the incoming matter can spin it up to millisecond 
periods over $10^8$~yrs..

In the standard theory of accretion disk,
there exists a radial point $r$ ($=r_{\rm iao}$, here defined as the 
innermost allowed orbit) in the disk where the gradient of 
the angular velocity profile changes its sign.  The accreting 
system, therefore, may be thought to comprise of (i) an extended accretion 
{\it disk}, which is a region extending from infinity inwards to 
$r_{\rm iao}$ and (ii) a {\it boundary layer}, a region between 
$r_{\rm iao}$ and the neutron star surface ($r=R$).  
In this talk, we examine how the incorporation of general
relativity, rotation, and the relevant interior physics of neutron stars 
into the calculations, re-order the gravitational energy release from 
the boundary layer, and the extended disk.

To calculate the gravitational energy release we first obtain the effective 
gravitational potential from the radial 
equation of motion for a material particle orbiting the central object. 
The conditions 
of circularity and extremisation of energy yield the specific energy 
($E(r)$) and specific angular momentum ($l(r)$) of particles in any orbit
of radius $r$.  
The gravitational energy release in the disk is given by 
$E_{\rm D}= E(\infty) - E(r_{\rm iao}$ and
that in the boundary layer is
$E_{\rm BL}=E(r_{\rm iao}) - E^{\ast}(R)$; with $E^{\ast}(R)$ defined 
as the specific energy of the particle at rest on the stellar surface.

In Newtonian approximation, $r_{\rm iao} = R$ (for all 
practical purposes) and the ratio: $E_{\rm BL}/E_{\rm D}= 1$.

\section{General Relativistic Effects}

\subsection{Structure: Equation of State (EOS)}
General relativity defines the structure of the compact
object.  
A key input to solve the structure (hydrostatic equilibrium) 
equation is the equation of state (EOS): 
$P(\rho)$ of high density matter.  For our purpose here, we use
three models of EOS:
(1) Bombaci (1995), BPAL12 (2) Wiringa, Fiks and Fabrocini (1988)
 UU (3) Sahu, Basu and Datta (1993).
These models, widely spaced in their qualitative properties, 
make the results sufficiently general.  

For Schwarzchild geometry, we solve the Tolman--Oppenheimer--Volkoff:
TOV equations.  For rapidly rotating 
neutron stars, taking the space--time geometry to be described by 
a general axisymmetric metric, we solve the 
hydrostatic equilibrium equations and Einstein equations 
numerically and self--consistently.  We construct rapidly rotating
configurations having rotation rates ranging from zero to centrifugal 
mass shed limit (e.g. Datta, Thampan \& Bombaci 1998).

\subsection{Equation of Motion}
For general relativistic effective 
potentials, the minimisation condition is satisfied only marginally,
implying the existence of a marginal stable orbit with $r=r_{\rm orb}$.  
Neutron stars, described by realistic EOS models, may have 
$R> r_{\rm orb}$  or $R< r_{\rm orb}$ (Sunyaev \& Shakura 1986;
Klu\'zniak \& Wagoner 1985). 
Accordingly, for accretion disks,
(i) $r_{\rm iao}=R$  for $R>r_{\rm orb}$ and
(ii) $r_{\rm iao}=r_{\rm orb}$ for $R<r_{\rm orb}$.

For non--rotating neutron stars, $E_{\rm BL}/E_{\rm D}$ can be as high
as 6 (Sunyaev \& Shakura 1986).  For neutron stars 
rotating at the mass shed limit, we obtain
$E_{\rm BL}/E_{\rm D} \sim 0$ (Thampan 2000).

These results are expected to be important for modeling 
accretion disks in LMXBs.

\section*{Acknowledgements}
The author thanks: the organizing committees MG9 for local
hospitality and the Indian Government Departments: DST and CSIR 
for travel support.  Late Professor Bhaskar Datta is thanked 
for his guidance and the talk is dedicated to his memory.

\end{document}